# Structure and vibrational properties of methane up to 71 GPa

Maxim Bykov[1,2], Elena Bykova[1], Chris J. Pickard[3,4], Miguel Martinez-Canales[5], Konstantin Glazyrin[6], Jesse S. Smith[7], Alexander F. Goncharov[1]

[1]Earth and Planets Laboratory, Carnegie Institution of Washington, 5251 Broad Branch Road Washington D.C., USA

[2]Howard University, 2400 6th St NW, Washington DC 20059, USA

[3]Department of Materials Sciences & Metallurgy, University of Cambridge, Cambridge, UK

[4]Advanced Institute for Materials Research, Tohoku University, Aoba, Sendai, 980-8577, Japan

[5]School of Physics & Astronomy, The University of Edinburgh, Edinburgh, UK

[6]Photon Sciences, Deutsches Electronen Synchrotron (DESY), D-22607 Hamburg, Germany

[7]HPCAT, X-ray Science Division, Argonne National Laboratory, Argonne, IL 60439, USA

**Single-crystal synchrotron X-ray diffraction, Raman spectroscopy, and first principles calculations have been used to identify the structure of the high-pressure (HP) phase of molecular methane above 20 GPa up to 71 GPa. The structure of HP phase is trigonal *R*3, which can be represented as a distortion of the cubic phase B, previously documented at 7-15 GPa and confirmed here. The positions of hydrogen atoms in HP phase have been obtained from first principles calculations. The molecules occupy four different crystallographic sites in phases B and eleven sites in the HP phase, which result in splitting of molecular stretching modes detected in Raman spectroscopy and assigned here based on a good agreement with the Raman spectra calculated from the first principles.**

## Introduction

Methane represents a rare and fundamental example of hydrogen-bearing molecular compound with no hydrogen bonds unlike other simple molecules HF, $H_2O$, $H_2S$, and $NH_3$. Thus, the crystal structure formation in methane solids is expected to be mainly determined by the steric effects and van der Waals bonds, suggesting that its behavior at high pressure may be similar to the rare gas solids, for example Ar [1], which persists in a face-centered cubic (*fcc*) structure to very high pressures. Furthermore, the high-pressure behavior of methane is of interest to explore possible deviations from simple closed packed structures and related to this molecular orientational ordering, where the hydrogen atoms find the most energetically favorable stable positions. The fate of methane at even higher compression states in the regime of molecular proximity (i.e. shortened intermolecular H-H distances) is of great interest as it can be considered as a hydrogen dominant alloy [2] with the relevance to high-temperature superconductivity [3]. In addition, the behavior of methane at high pressures and temperatures may be of interest because it is believed to be a valued component of the deep planetary interiors of "icy" giant planets such as Uranus and Neptune [4]. Such hot molecular ices of water and ammonia form superionic crystals with diffusive protons at extreme pressure-temperature (P-T) conditions of relevance for these planets [5-7], while methane tends to chemically react yielding oligomeric hydrocarbon compounds [8] and dissociate at higher P-T conditions yielding diamond and hydrogen [9-11].

Similar to Ar, methane crystallizes in phase I at about 1.5 GPa upon compression at room temperate forming a face-centered cubic (*fcc*) closed packed structure of the molecular centers at the carbon atoms [12], and this crystal is plastic - molecules are freely rotating. However, unlike in noble gases, this structure transforms at 5.4 GPa to a complex rhombohedral structure A with 21 molecules in the unit cell solved by a combination of single crystal neutron and X-ray diffraction (XRD) techniques [13]. The structure is rhombohedrally distorted from the *fcc* methane I, with the carbon atoms occupying nine different crystallographic positions, and some of them may be associated with orientationally ordered molecules. However, the Raman and IR spectra do not show the obvious splitting of the major $\nu_1$ and $\nu_3$ C-H stretching modes [14].

Further compression leads to a transition to phase B with a very large unit cell containing 58 molecules, which has been found cubic as solved recently by Maynard-Casely *et al*. [15] for the carbon subsystem by combining powder and single-crystal XRD. The phase B shows a splitting of $\nu_1$ and $\nu_3$ stretching C-H modes in IR and Raman spectra [1, 14, 16, 17], suggesting that it possesses some elements of orientational order (*e.g*. Refs. [14, 16]). The transition from phase A to phase B at about 8 GPa is very sluggish and phase B can be missed if fast compressed yielding a simple cubic phase pre-B [16], where the Raman active $\nu_1$ and $\nu_3$ C-H stretching modes are not split suggesting that methane molecules are orientationally disordered. Phase B is reported to transform to a high pressure phase 1 (HP1) at about 25 GPa manifested by anomalies in IR spectra, vibrational frequencies [18], and specific volume [19], while no drastic structural changes were detected in the powder diffraction patterns. At higher pressures, Raman spectroscopy observations suggest further structural modifications to HP2 and HP3 phases reported up to 62 GPa; however, the transition pressures to the subsequent high-pressure phases HP2 and HP3 are contradictory [16, 17]. The structure of the HP phase(s) has not been solved; it has been postulated to be cubic (*e.g.* Ref. [20]), but no detailed structural information is available. The XRD experiments have been extended to 202 GPa and the volume has been determined based on a cubic symmetry, even though a major transformation has been detected and a volume discontinuity was found at 94 GPa [20]. Methane was reported to persist in a molecular crystal structure beyond 200 GPa [20, 21].

In this paper, we examine the structure of methane in the pressure range between 7 and 71 GPa using single-crystal XRD, Raman spectroscopy, and first-principles theoretical calculations. Our single-crystal XRD data directly determine the structure of the carbon subsystem in agreement with the previous report in phase B [15], while we find that HP phase can be viewed as rhombohedrally distorted phase B (space group *R*3) with 87 molecules in the unit cell (29 in the primitive unit cell). We applied *ab initio* random structure search [22] of the most stable structures with and without any restriction of the symmetry. The results yield the structure of HP phase in agreement with XRD data, and determine the hydrogen positions in HP phase. The Raman spectrum of this phase calculated from the first principles is in a qualitative agreement with the experiment demonstrating that the C-H stretching mode splitting is due to a variation in the crystal field at different crystallographic sites.

**Experimental and theoretical procedures**

Methane gas was loaded in a mixture with He gas at approximately 1:1 mole ratio compressed to 0.15 GPa at room temperature in a BX90 diamond anvil cell [23] (Fig. S1 of the Supplemental Material [24]). In order to obtain good-quality single crystals of phase B of methane, the gas mixture was slowly compressed at 297 K through the methane solidification point at 1.5 GPa and then to 8 GPa. At this pressure, excellent single crystals of phase B were grown overnight judging on Raman observations of a splitting of the $\nu_1$ C-H stretch mode (Fig. S2 of the Supplemental Material [24]). These crystals yield very sharp Bragg reflections with typical rocking-curve width of 0.64° at 7.1 GPa (Fig. 1). Two samples (#1 and #2) were compressed up to 39 GPa and 71 GPa respectively and concomitant XRD and Raman spectroscopy measurements (Figs. 1, 2 and Fig. S2 of the Supplemental Material [24]) were taken at selected pressure points. All the experiments were performed at room temperature (295(2) K).

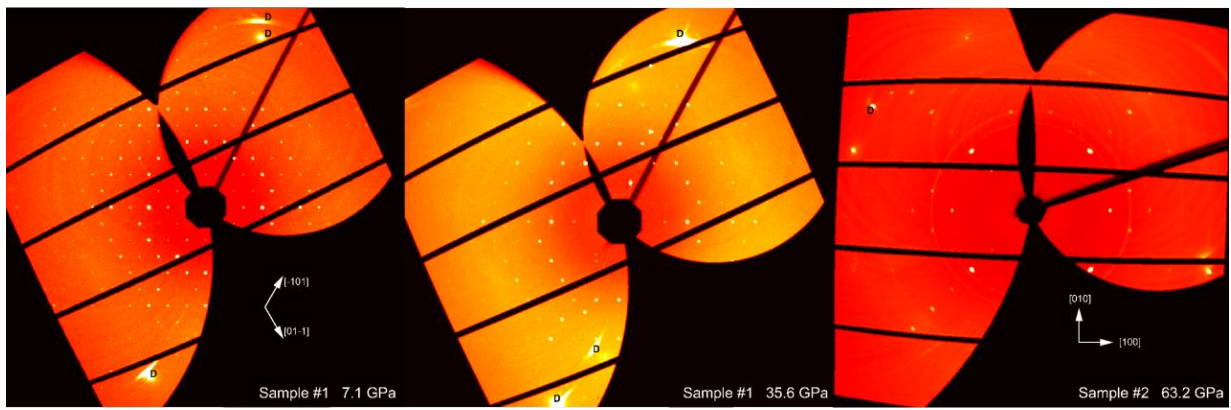

**Figure 1**. Reconstructed reciprocal lattice plane of methane visualized by CrysAlisPro software package[25] at 7.07 GPa (left), 35.6 GPa (middle) and 63.2 GPa (right). **D** stands for diamond reflections. The observed diffraction spots from the sample have been indexed and used to determine the structure of phases B (left panel) and high-pressure (HP) phase (middle and right panel) (Table S1 and Tables 1, 2, respectively).

X-ray diffraction was performed at the beamlines 16-ID-B (APS, Argonne, USA) and P02.2 (Petra III, DESY, Hamburg, Germany). The following beamline setups were used. P02.2: $\lambda$ = 0.289 Å, beam size ~2×2 μm$^2$, Perkin Elmer XRD 1621 detector; 16-ID-B: $\lambda$ = 0.344 Å, beam size ~3×3 μm$^2$, Pilatus 1M detector. For the single-crystal XRD measurements samples were rotated around a vertical ω-axis in a range ±22°. The diffraction images were collected with an angular step Δω = 0.5° and an exposure time of 2-10 s/frame. For analysis of the single-crystal diffraction data (indexing, data integration, frame scaling and absorption correction) we used the *CrysAlis$^{Pro}$* software package[25]. To calibrate an instrumental model in the *CrysAlis$^{Pro}$* software, *i.e.*, the sample-to-detector distance, detector's origin, offsets of goniometer angles, and rotation of both X-ray beam and the detector around the instrument axis, we used a single crystal of orthoenstatite (($Mg_{1.93}Fe_{0.06}$)($Si_{1.93}$, $Al_{0.06}$)$O_6$, *Pbca* space group, $a$ = 8.8117(2), $b$ = 5.18320(10), and $c$ = 18.2391(3) Å). The same calibration crystal was used at both beamlines. The structure was solved with the ShelXT structure solution program[26] using intrinsic phasing and refined with the Jana 2006 program [27]. Raman spectra were collected using the GSECARS Raman System[28] with the excitation wavelength of 532 nm and at the Earth and Planets Laboratory of Carnegie (EPL) using a 488 nm excitation line (e.g. Ref. [29]). Pressure was determined based on the shift of the R1 ruby fluorescence line [30].

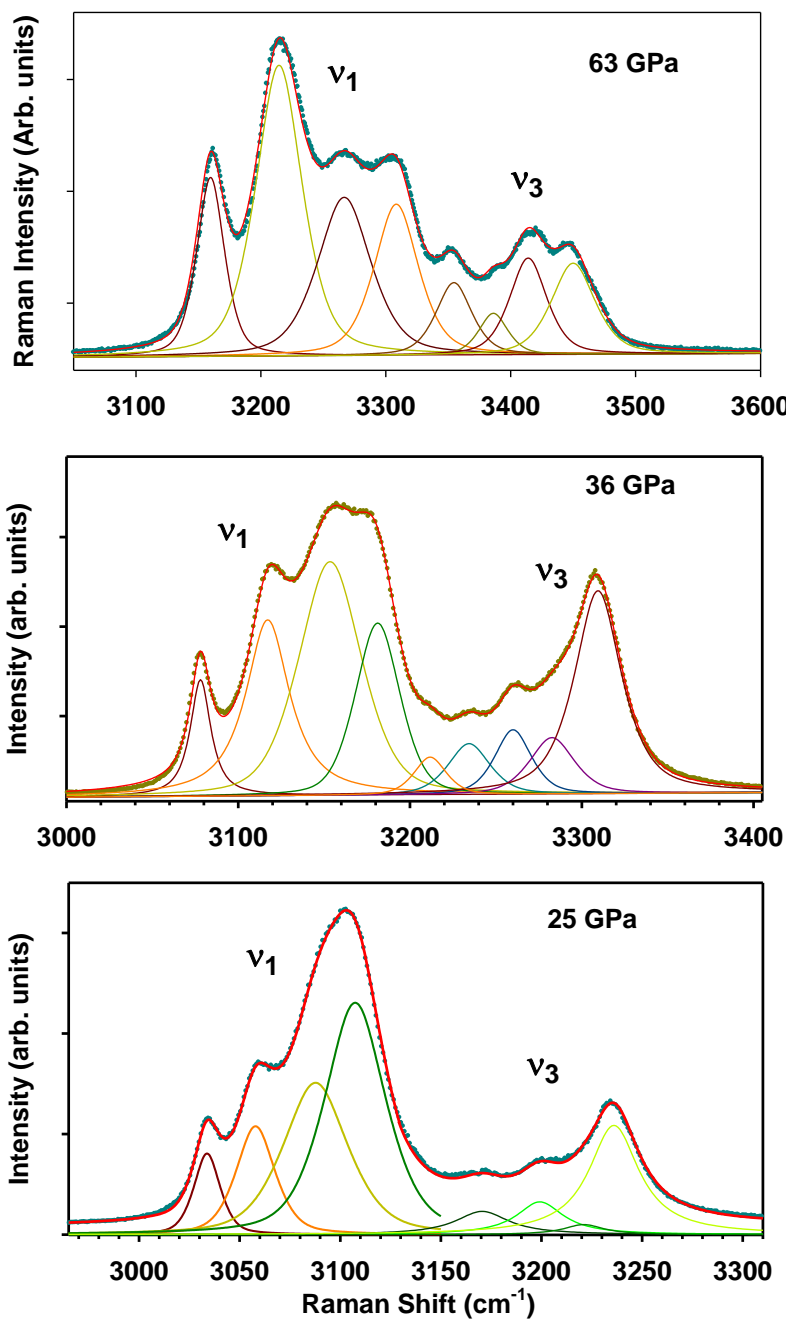

**Figure 2. Raman spectra of the C-H stretching modes of methane in HP phase at various pressures.** The dots are the data and the lines are the fits to the Voigt oscillator model with eight peaks; variable color lines show the individual $\nu_1$ and $\nu_3$ components.

Ab Initio Random Structures Searches (AIRSS) [31] were performed using the CASTEP code [22] to explore the structures of the HP methane with the rhombohedral symmetry (the unit cells with up to 87 atoms were explored), with a subsequent relaxation using density functional theory (DFT) calculations within the CASTEP code [22]. Raman spectra were computed using density functional perturbation theory (DFPT) + Finite differences method, using the default CASTEP 18 NC pseudopotentials with a 1250 eV energy cutoff, and a similarly dense k-point grid.

**Results and discussion**

At pressures below ~16 GPa, the diffraction pattern of $CH_4$ (Fig. 1) can be indexed with a body-centered cubic unit cell of phase B (Table S1 and Fig. S3 [24]). Structure solution and refinement is in agreement with the model reported by Maynard-Casely *et al.* [15]. Due to weak scattering of hydrogen atoms, we were able to refine the positions of carbon atoms only. The unit cell contains four independent atomic sites: C1, C2, C3, and C4 occupying the positions 2$a$, 24$g$, 8$c$ and 24$g$, respectively. Therefore, there are 58 $CH_4$ molecules in the unit cell.

Further compression leads to a noticeable lattice distortion at above 18 GPa and the unit cell can no longer be considered cubic. This distortion is accompanied by the appearance of new modes in the Raman spectrum (Figs. 2 and Fig. S2 of the Supplemental Material [24]) in agreement with the previous observations [16, 17] at similar pressure conditions. The most plausible indexing of the diffraction patterns of a methane high-pressure phase (HP) above 18 GPa can be performed using a rhombohedral unit cell (Tables 1, 2). Symmetry-lowering transitions from cubic *I*-43$m$ to rhombohedral symmetry may follow $\Gamma_4$ or $\Gamma_5$ irreducible representations with the following transformation of the lattice parameters:

$$\begin{pmatrix} a_R \\ b_R \\ c_R \end{pmatrix} = \begin{pmatrix} -1 & 1 & 0 \\ 0 & -1 & 1 \\ 1/2 & 1/2 & 1/2 \end{pmatrix} \begin{pmatrix} a_c \\ b_c \\ c_c \end{pmatrix},$$

where $a_R$, $b_R$, $c_R$ are lattice vectors of the rhombohedral unit cell (hexagonal setting), while $a_c$, $b_c$, $c_c$- lattice vectors of the cubic unit cell. The $\Gamma_4$ distortion leads to the space group symmetry *R*3$m$ (#160), while $\Gamma_5$ - to *R*3 (#146). We should note that at pressures above 30 GPa we see a slight departure of the lattice parameters from *R*-centered hexagonal lattice ($a = b$, $\alpha = \beta = 90$, $\gamma = 120°$) suggesting a further monoclinic distortion. However, the degree of this distortion is not reproducible between two experiments (Fig. S4). Therefore, we conclude that this effect is related to the development of non-hydrostatic stresses in the sample, which is unavoidable at very high pressures and especially for very soft crystals like methane. We therefore treat the high-pressure phase of methane as a single $R3$ phase.

We have tested both $R3$ and $R3m$ models (Table 1). Although the $R3$ model has smaller agreement factors, they are not significantly different from those of the $R3m$ model. We have performed a Hamilton significance test in order to check if the improvement of the agreement factors for the $R3$ model may be considered significant [32]. For this test, both models were refined using the same set of reflections averaged based on the $R3$ symmetry. Based on the Hamilton test, the $R3$ model is preferable with the 75-90 % confidence level depending on the pressure point. We, therefore, cannot completely reject the $R3m$ model based on the XRD data, but can conclude that the $R3$ model is more likely (Table 1).

In order to complete the structure solution and determine the positions of hydrogen atoms, theoretical structure optimizations have been performed at various pressures (Fig. 3). First, we performed ab initio Random Structure Search (AIRSS) using the CASTEP code [22]. All calculations used the PBE functional together with the DFT-D dispersion correction scheme of Tkatchenko and Scheffler [33]. Among a plethora of phases marginally different in enthalpies, two structures emerged having $R3$ symmetries and 29 molecules in the primitive unit cell ($R3$-I and $R3$-II on the Fig. 2(a)) in agreement with the experimental HP phase (Fig. 4). Both calculated $R3$ structures have equivalent carbon substructures and slightly different orientations of $CH_4$ molecules centered at the C4 atoms. According to the theoretical calculations both $R3$ structures are not the most energetically favorable at 0 K. However, if temperature effects are taken into account, the calculated Gibbs free energies show that the $R3$-II phase becomes the most favorable above 260 K at 25 GPa, which perfectly agrees with the experiment (Figure 3(b)). A monoclinic distortion noticed experimentally is found to be energetically unfavorable. Below we will mainly discuss the most energetically favorable $R3$ phase ($R3$-II, Table 3).

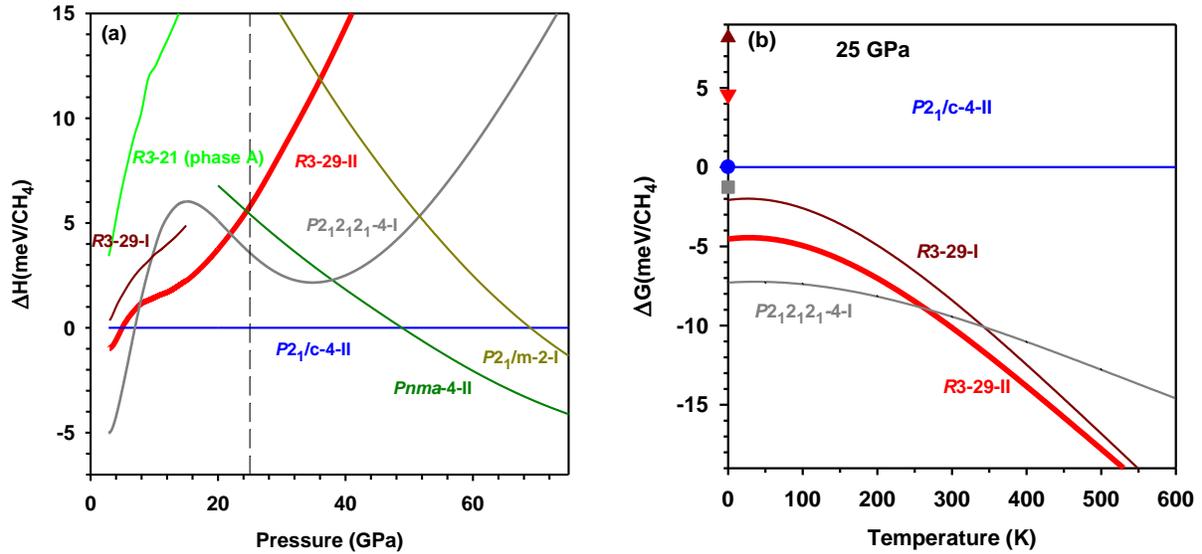

**Figure 3**. (a) Theoretically calculated relative enthalpy of the most stable methane phases. The results for HP phase ($R3$-II) are shown by a thick red line. Other orientationally ordered phases are shown by thin lines of various colors. The vertical dashed line correspond to the pressure at which the temperature effects were taken into account (b) Relative Gibbs free energies of the most stable methane phases as a function of temperature at 25 GPa. The symbols at T=0 K represent relative classical enthalpies as computed with norm-conserving pseudo potentials.

The unit cell of HP-$CH_4$ (Fig. 4 and Fig. S5 of the Supplemental Material [24]) contains 11 symmetry independent carbon atoms with atoms C1_1 and C3_1 occupying positions 3*a*, while the rest occupying positions 9*b*. Here, the C atoms are named following the numbering for phase B [15] to show the relation between B and HP phases, *i.e.* the atom C3_1 is generated from the atom C3 of the B-phase.

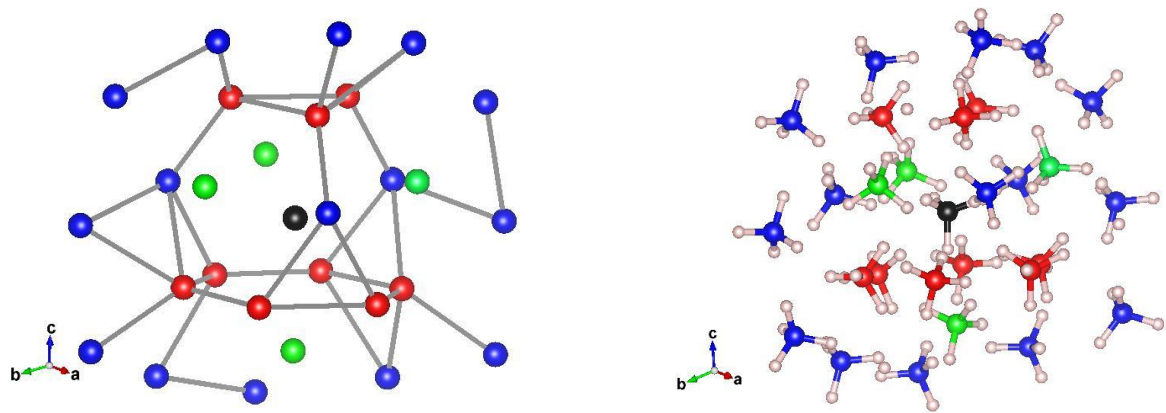

**Figure 4**. Crystal structure of HP-CH$_4$ at 35 GPa. Black, red, green and blue balls represent the positions of C1, C2, C3, and C4 carbon atoms, respectively; small pinkish circles are hydrogen atoms. Left panel shows C atoms only; gray lines connect the nearest C atoms with a cutoff of 3.08 Å.

Figure 4 shows the main motif of the most energetically favorable *R*3 structure. The C1_1 and C3_1 atoms that lie on the three-fold axis possess a distorted tetrahedral icosi-octahedral environment. This polyhedron can be understood in the following way (see bottom of the Fig. S5 [24]): the atom C3_1 is coordinated by two six-membered rings, one of which is planar and capped, while the other is in a chair arrangement capped by a triangle. The packing of HP-CH$_4$ is primarily defined by intermolecular H-H interactions. The histogram of H-H distances has two distinct peaks (Fig. S6 [24]). The sharp peak at ~1.75 Å corresponds to intramolecular H-H distances and almost does not shift with pressure as the molecules remain almost rigid under compression. The broad maxima at ~2.05 and ~1.95 Å at 20 and 30 GPa, respectively, correspond to closest intermolecular H…H contacts. These distances are in a good agreement with shortest intramolecular H…H contacts in compressed hydrogen [34].

The $\nu_1$ and $\nu_3$ C-H stretching Raman modes become composite (see also Refs. [16, 17]) in phases B and HP suggesting either site symmetry or vibrational splitting. The Raman spectra collected at 25 GPa in HP phase show that each of the $\nu_1$ and $\nu_3$ multifold can be well represented by 4 peaks (Figs. 2, 5). On the other hand, theoretical calculations of the Raman activity of HP phase performed in this work show a larger number of components, which correspond to various molecular sites as the analysis for the $\nu_1$ modes shows (Fig. 5). However, the splitting pattern and value are similar to the experimental; moreover, some theoretically calculated peaks are very close to each other making them difficult to resolve in the experiment; this likely results in broadened peaks, which consist of many components. A prominent feature of the $\nu_1$ Raman spectra is a well split off low frequency peak, that theory predicts is due internal vibrations of molecules involving C1 and C3 carbon in 3*a* sites, which are characterized by relatively more uniform in length and overall longer intermolecular distances (Fig. S7 [24]). This distinction results in a well-separated low-frequency $\nu_1$ component, which can be also seen in phase B with similar carbon site geometry save the lattice distortion.

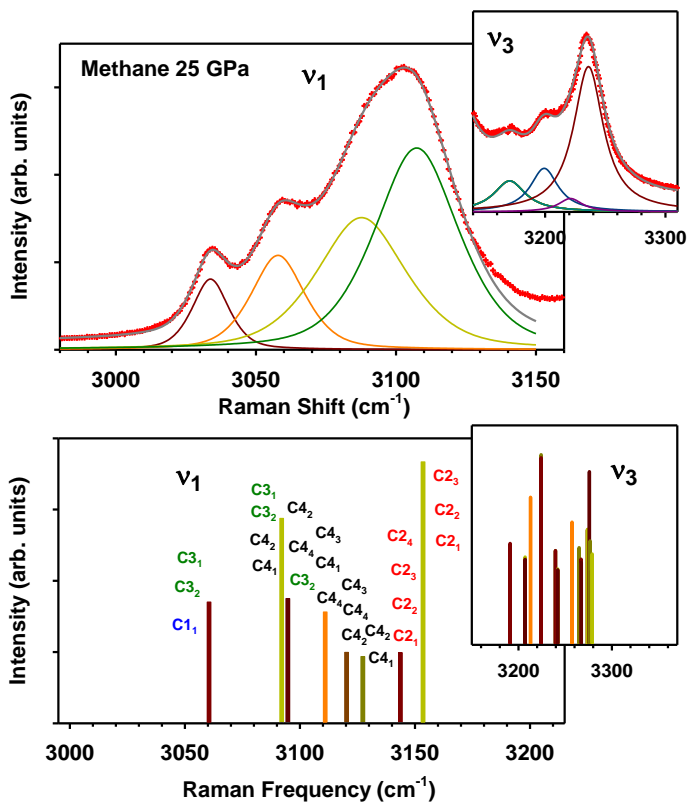

**Figure 5**. Raman spectra of $\nu_1$ (main panel) and $\nu_3$ (insets) stretching modes. The top panels is the spectrum measured at 25 GPa (488 nm excitation). Crosses are the data and gray line is the best fit to the Voigt oscillator model; variable color lines show the individual $\nu_1$ and $\nu_3$ components. The bottom panel shows the theoretically computed spectra at the same pressure. The labels (carbon atom sites) show the assignment of the peaks to the molecules of different kind. The frequency axes in the top and the bottom panels are scaled for the best correspondence of the experimental and theoretical peaks.

The unit-cell parameters and symmetry of phases B and HP obtained here experimentally and theoretically provide reliable information on the P-V equation of state of methane (Fig. 6). Our results agree well with the previous determinations based on single-crystal [13, 15] and powder [19, 20, 35] XRD. The cubic unit cell of phase B shows a distortion in the HP phase, which first appears small but then increases with pressure (Fig. S4 [24]). Please note that pre-B phase, which appears on fast compression of phase A, was indexed as a simple cubic lattice [19, 20]; this phase is not examined with XRD here (but was detected by Raman spectroscopy to coexist with HP phase in the experiment #2 up to the highest pressure investigated) but it is clearly different from HP phase of this study. HP and pre-B methane phases have very close volumes per the formula unit (Fig. 6) if one assumes that pre-B has 21 molecules in the unit cell as in phase A. Unlike previously reported modifications of HP phase based on changes in a number of $\nu_1$ and $\nu_3$ Raman components [16, 17], our direct structural data (Fig. 1, Tables 1-2) and analysis point out into the sole HP phase up to 71 GPa (cf. Refs. [16, 17]), where the rhombohedral distortion gradually increases with pressure (Fig. S4 [24]). Consistently, we also find (by fitting the spectra using an oscillator models, Figs. 2, 5) that the number of C-H Raman components remains the same (cf. Refs. [16, 17]) and their frequencies (Fig. 7 [24]) agree well with the results of previous Raman investigations [16, 17]. Previous Raman investigations [16, 17] reported minor changes in the spectra (peak splitting and changes in peak intensities) and in the P dependence of the C-H stretch frequencies. However, these changes do not signify the phase transitions. The Raman peaks may split because they consist of a number of components, which have different pressure slopes so they become separated with pressure, while their relative intensity may change due to a number of reasons including the mode coupling and modification in the orientational order.

The unit cell volume changes smoothly through the B-HP transition suggesting a very small or no discontinuity. The theoretically computed lattice parameters and volumes are smaller than those determined theoretically, but the discrepancy is within 3%, which commonly is considered acceptable. However, this small difference may indicate that HP methane may still be partially disordered, which also corroborate with smaller (by some 20%) observed values of $\nu_1$ and $\nu_3$ multifolds splitting compared to the computed ones (Fig. 5).

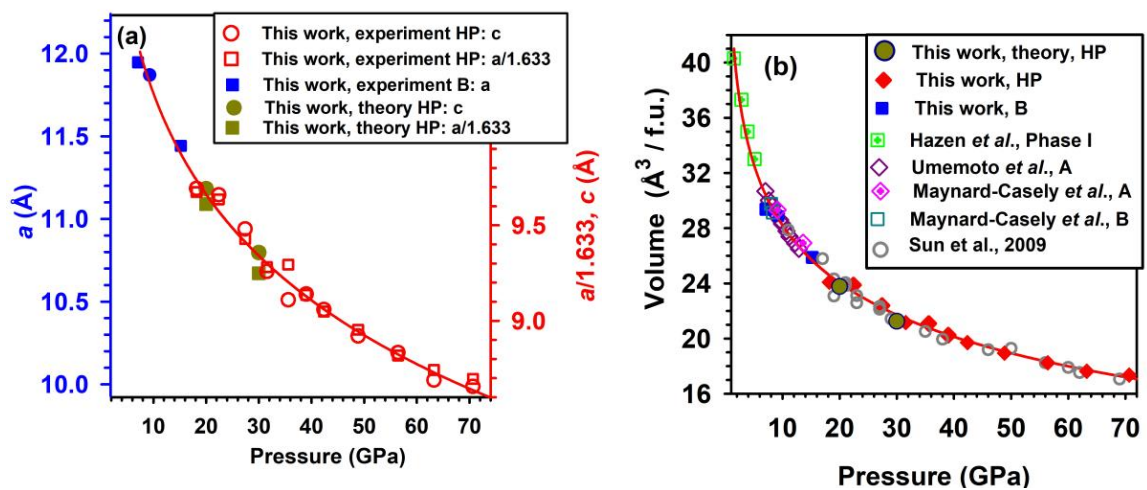

**Figure 6**. Lattice parameters (a) and volumes per formula unit (b) of methane as a function of pressure. Note two vertical scales in the panel (a) for phases B (blue) and HP (red) which are scaled by a factor of $\sqrt{3}/2$ reflecting the difference in the unit cell dimensions. Also, the parameter *a* of the HP phase is divided by a factor of 1.633. Our experimental and theoretical volumes (b) are compared to previous experiments in phases I, A, B, and HP [12, 13, 15, 19, 20]. The unit cell volume of Ref. [20] was assumed to contain 21 formula units. The lines are guides to the eye.

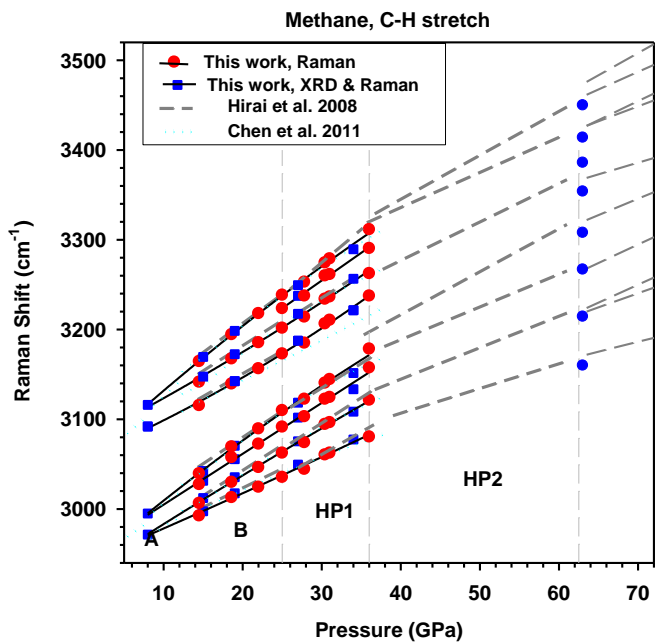

**Figure 7. Raman frequencies of the C-H stretching modes of methane as a function of pressure.** Vertical dashed lines mark the phase transitions following Ref. [16]. The symbols are the results of two experiments of this work (solid line is the guide to the eye), while dashed and dotted lines represent the results of previously published works [16, 17].

**Conclusions**

Our concerted experimental and theoretical investigation determined the structure of the high-pressure HP phase of methane including the positions of hydrogen atoms. The structure is rhombohedral (space group *R*3) with 87 molecules in the unit cell. This extraordinary complex structure of approximately spherical molecules is in a drastic contrast with simple *fcc (hcp)* structures of noble gas solids. Moreover, the theoretically predicted orientationally ordered structures with smaller unit cell and with well elaborated intermolecular coupling schemes also do not capture the real structure. This indicates that the observed structure represents a balance between enthalpy and entropy terms thus suggesting that some molecular sites remain orientationally disordered; this makes this structure analogous to clathrates, where host and guest molecules are the same.

Our direct single-crystal XRD results suggest that there is only one high-pressure phase (HP) of methane between 20 and 71 GPa in contrast to previous works, which suggested three HP phase based on the results of Raman spectroscopy. HP phase investigated here can coexist with a cubic pre-B phase investigated previously, the detailed structure of which still remains unknown.

**Acknowledgements:**

Parts of this research were carried out at the Extreme Conditions Beamline (P02.2) at DESY, a member of Helmholtz Association (HGF). Portions of this work were performed at HPCAT (sector 16) of the Advanced Photon Source (APS), Argonne National Laboratory. HPCAT operations are supported by DOE-NNSA's Office of Experimental Sciences. Concomitant Raman spectroscopy experiments were performed at GeoSoilEnviroCARS (The University of Chicago, Sector 13), Advanced Photon Source (APS), Argonne National Laboratory. GeoSoilEnviroCARS is supported by the National Science Foundation - Earth Sciences (EAR - 1634415) and Department of Energy GeoSciences (DE-FG02-94ER14466). The Advanced Photon Source is U.S. Department of Energy (DOE) Office of Science User Facility operated for the DOE Office of Science by Argonne National Laboratory under Contract No. DE-AC02-06CH11357. CJP acknowledges financial support from the Engineering and Physical Sciences Research Council [Grant EP/P022596/1]. MB research was sponsored by the Army Research Office and was accomplished under the Cooperative Agreement Number W911NF-19-2-0172.

**Table 1**. Comparison of agreement factors between $R3$ and $R3m$ models of the HP methane

| Pressure, GPa | # reflections | $R3$ model | | $R3m$ model | |
|---|---|---|---|---|---|
| | | # parameters | $R_1$ | # parameters | $R_1$ |
| 22.4 | 381 | 41 | 0.127 | 28 | 0.1303 |
| 27.4 | 474 | 41 | 0.138 | 28 | 0.1422 |
| 35.6 | 280 | 41 | 0.1020 | 28 | 0.1033 |
| 39.0 | 198 | 41 | 0.1415 | 28 | 0.1436 |

**Table 2.** Crystal data, refinement and crystal structure details of Methane HP phase

| Pressure (GPa) | 22.4 | 27.4 | 35.6 | 39.0 |
|---|---|---|---|---|
| **Crystal data** | | | | |
| Crystal system, space group | trigonal, $R3$ | | | |
| Z | 87 | | | |
| $a$, $c$ (Å) | 15.735(3), 9.660(2) | 15.396(3), 9.481(2) | 15.176 (8), 9.11 (7) | 14.918(8), 9.14(7) |
| $V$ (Å$^3$) | 2071.3(7) | 1946.1(7) | 1818(13) | 1762(14) |
| Radiation type | X-ray, $\lambda = 0.34453$ Å | | | X-ray, $\lambda = 0.2891$ Å |
| $\mu$ (mm$^{-1}$) | 0.02 | 0.02 | 0.02 | 0.02 |
| **Data collection** | | | | |
| Diffractometer | GP @ ID16-B | GP @ ID16-B | GP @ ID16-B | GP @ P02.2 |
| Absorption correction | Multi-scan | | | |
| No. of measured, independent and observed [$I > 3\sigma(I)$] reflections | 676, 663, 381 | 640, 627, 474 | 475, 365, 280 | 792, 587, 198 |
| $R_{int}$ | 0.111 | 0.026 | 0.039 | 0.059 |
| $(\sin \theta/\lambda)_{max}$ (Å$^{-1}$) | 0.752 | 0.768 | 0.586 | 0.664 |
| **Refinement** | | | | |
| $R[F^2 > 2\sigma(F^2)]$, | 0.127, 0.136, | 0.138, 0.150, | 0.102, 0.117, | 0.141, 0.149, |

| $wR(F^2)$, $S$ | 3.14 | 5.70 | 3.61 | 1.60 |
|---|---|---|---|---|
| No. of reflections | 663 | 627 | 365 | 587 |
| $\Delta\rho_{max}$, $\Delta\rho_{min}$ (e Å$^{-3}$) | 0.32/-0.28 | 0.36, -0.37 | 0.25, -0.19 | 0.23, -0.20 |
| **Crystal structure** | | | | |
| **C1_1** | | | | |
| x | 0 | 0 | 0 | 0 |
| y | 0 | 0 | 0 | 0 |
| z | 0.02(12) | 0.02(8) | 0.000(17) | 0.013(9) |
| U$_{iso}$ | 0.09(11) | 0.090(12) | 0.049(8) | 0.033(4) |
| **C2_1** | | | | |
| x | 0.0618(8) | 0.0644(9) | 0.0623(8) | 0.0635(9) |
| y | 0.1264(8) | 0.1264(9) | 0.1234(9) | 0.1251(8) |
| z | 0.33(12) | 0.33(8) | 0.340(16) | 0.340(8) |
| U$_{iso}$ | 0.045(3) | 0.035(3) | 0.037(4) | 0.050(3) |
| **C2_2** | | | | |
| x | 0.4816(9) | 0.4801(9) | 0.4844(8) | 0.4863(8) |
| y | 0.1468(9) | 0.1451(9) | 0.1485(8) | 0.1476(7) |
| z | 0.17(12) | 0.17(8) | 0.165(15) | 0.174(8) |
| U$_{iso}$ | 0.041(3) | 0.028(3) | 0.028(3) | 0.046(3) |
| **C2_3** | | | | |
| x | 0.6663(9) | 0.6653(9) | 0.6632(7) | 0.6622(7) |
| y | 0.1458(9) | 0.1477(9) | 0.1483(7) | 0.1495(7) |
| z | 0.17(12) | 0.17(8) | 0.162(16) | 0.166(8) |
| U$_{iso}$ | 0.039(3) | 0.028(3) | 0.029(3) | 0.041(3) |
| **C2_4** | | | | |
| x | 0.1216(7) | 0.1230(8) | 0.1256(7) | 0.1245(7) |
| y | 0.2456(7) | 0.2449(8) | 0.2489(6) | 0.2506(7) |
| z | 0.09(12) | 0.09(8) | 0.101(16) | 0.093(9) |
| U$_{iso}$ | 0.041(3) | 0.035(3) | 0.034(3) | 0.046(3) |
| **C3_1** | | | | |
| x | 0 | 0 | 0 | 0 |
| y | 0 | 0 | 0 | 0 |
| z | 0.67(12) | 0.67(8) | 0.666(16) | 0.670(10) |
| U$_{iso}$ | 0.016(4) | 0.014(4) | 0.015(4) | 0.049(6) |
| **C3_2** | | | | |
| x | 0.2388(6) | 0.2376(7) | 0.2361(6) | 0.2339(7) |
| y | 0.1208(6) | 0.1174(7) | 0.1178(6) | 0.1175(7) |
| z | 0.14(12) | 0.14(8) | 0.152(15) | 0.156(8) |
| U$_{iso}$ | 0.018(2) | 0.014(2) | 0.019(3) | 0.039(3) |
| **C4_1** | | | | |
| x | 0.4644(8) | 0.4595(8) | 0.4648(7) | 0.4635(7) |
| y | -0.0756(7) | -0.0774(8) | -0.0704(6) | -0.0730(7) |
| z | 0.24(12) | 0.24(8) | 0.244(16) | 0.239(8) |
| U$_{iso}$ | 0.047(3) | 0.038(3) | 0.036(3) | 0.047(3) |
| **C4_2** | | | | |
| x | 0.0092(9) | 0.0123(10) | 0.0065(9) | 0.0062(7) |
| y | 0.3087(10) | 0.3130(11) | 0.3133(9) | 0.3104(7) |
| z | 0.33(12) | 0.33(8) | 0.339(16) | 0.337(8) |
| $U_{iso}$ | 0.050(4) | 0.049(4) | 0.036(3) | 0.038(3) |
| **C4_3** | | | | |
| x | 0.3005(8) | 0.2991(10) | 0.3080(9) | 0.3070(7) |
| y | 0.3121(9) | 0.3081(11) | 0.3116(9) | 0.3122(7) |
| z | 0.33(12) | 0.33(8) | 0.350(16) | 0.342(8) |
| $U_{iso}$ | 0.047(4) | 0.043(4) | 0.037(3) | 0.038(3) |
| **C4_4** | | | | |
| x | 0.2294(8) | 0.2239(10) | 0.2251(8) | 0.2230(8) |
| y | 0.4529(8) | 0.4531(10) | 0.4496(9) | 0.4481(8) |
| z | 0.19(12) | 0.18(8) | 0.166(15) | 0.154(8) |
| $U_{iso}$ | 0.038(3) | 0.039(3) | 0.030(3) | 0.040(3) |

**Table 3**. Crystal structure of HP-CH$_4$ at 25 GPa optimized by calculations ($R3$, $a$ = 15.3315, $c$ = 9.5221 Å).

| Site | $x$ | $y$ | $z$ |
|---|---|---|---|
| C1_1 | 0 | 0 | -0.9996 |
| H1_1_1 | -0.365771 | -0.622456 | -0.628642 |
| H1_1_2 | 0 | 0 | -0.11349 |
| C2_1 | -0.131193 | -0.066507 | -0.68994 |
| H2_1_1 | -0.181504 | -0.146561 | -0.695142 |
| H2_1_2 | -0.083784 | -0.042373 | -0.781794 |
| H2_1_3 | -0.085473 | -0.048146 | -0.596947 |
| H2_1_4 | -0.174924 | -0.029387 | -0.685741 |
| C2_2 | 0.148245 | 0.474857 | -0.519002 |
| H2_2_1 | 0.079276 | 0.476659 | -0.509461 |
| H2_2_2 | 0.205456 | 0.534268 | -0.456087 |
| H2_2_3 | 0.136305 | 0.402921 | -0.484066 |
| H2_2_4 | -0.162001 | -0.181677 | -0.294102 |
| C2_3 | -0.183006 | 0.005534 | -0.191981 |
| H2_3_1 | -0.255161 | -0.062329 | -0.199231 |
| H2_3_2 | 0.144078 | -0.273369 | -0.461258 |
| H2_3_3 | -0.129179 | -0.010913 | -0.144596 |
| H2_3_4 | -0.157189 | 0.036738 | -0.29511 |
| C2_4 | -0.252732 | -0.13696 | -0.448695 |
| H2_4_1 | -0.285273 | -0.116555 | -0.534236 |
| H2_4_2 | -0.300828 | -0.15757 | -0.357472 |
| H2_4_3 | -0.243439 | -0.199203 | -0.480598 |
| H2_4_4 | -0.180176 | -0.073615 | -0.422955 |
| C3_1 | 0 | 0 | -0.346612 |
| H3_1_1 | 0.044147 | -0.032218 | -0.309195 |
| H3_1_2 | 0 | 0 | -0.460331 |
| C3_2 | -0.096727 | -0.545671 | -0.549019 |
| H3_2_1 | 0.173063 | 0.094993 | -0.812157 |
| H3_2_2 | -0.079162 | -0.473773 | -0.593158 |
| H3_2_3 | -0.447561 | -0.266257 | -0.299235 |
| H3_2_4 | 0.299905 | 0.127484 | -0.823554 |
| C4_1 | -0.207481 | -0.422449 | -0.598684 |
| H4_1_1 | -0.274275 | -0.422628 | -0.563519 |
| H4_1_2 | -0.226509 | -0.496916 | -0.631667 |
| H4_1_3 | -0.173923 | -0.37107 | -0.685189 |
| H4_1_4 | -0.154241 | -0.399094 | -0.513821 |
| C4_2 | -0.322138 | -0.356324 | -0.361957 |
| H4_2_1 | -0.047361 | 0.238783 | -0.656792 |
| H4_2_2 | 0.083251 | 0.322741 | -0.659402 |
| H4_2_3 | -0.33452 | -0.297433 | -0.322463 |
| H4_2_4 | -0.323976 | -0.355022 | -0.47503 |
| C4_3 | -0.316752 | -0.01339 | -0.682065 |
| H4_3_1 | -0.308246 | 0.013518 | -0.575783 |
| H4_3_2 | 0.281839 | 0.313491 | -0.391824 |
| H4_3_3 | -0.252204 | 0.038722 | -0.743068 |
| H4_3_4 | 0.342394 | 0.246027 | -0.349393 |
| C4_4 | 0.214413 | 0.114945 | -0.509708 |
| H4_4_1 | 0.228418 | 0.190425 | -0.496222 |
| H4_4_2 | 0.141492 | 0.062025 | -0.467591 |
| H4_4_3 | 0.272312 | 0.106787 | -0.457289 |
| H4_4_4 | 0.214746 | 0.101117 | -0.620706 |

# Supplemental Material

# Structure and vibrational properties of methane up to 71 GPa

Maxim Bykov[1,2], Elena Bykova[1], Chris J. Pickard[3,4], Miguel Martinez-Canales[5], Konstantin Glazyrin[6], Jesse S. Smith[7], Alexander F. Goncharov[1]

[1]Earth and Planets Laboratory, Carnegie Institution of Washington, 5251 Broad Branch Road Washington D.C., USA

[2]Howard University, 2400 6th St NW, Washington DC 20059, USA

[3]Department of Materials Sciences & Metallurgy, University of Cambridge, Cambridge, UK

[4]Advanced Institute for Materials Research, Tohoku University, Aoba, Sendai, 980-8577, Japan

[5]School of Physics & Astronomy, The University of Edinburgh, Edinburgh, UK

[6]Photon Sciences, Deutsches Electronen Synchrotron (DESY), D-22607 Hamburg, Germany

[7]HPCAT, X-ray Science Division, Argonne National Laboratory, Argonne, IL 60439, USA

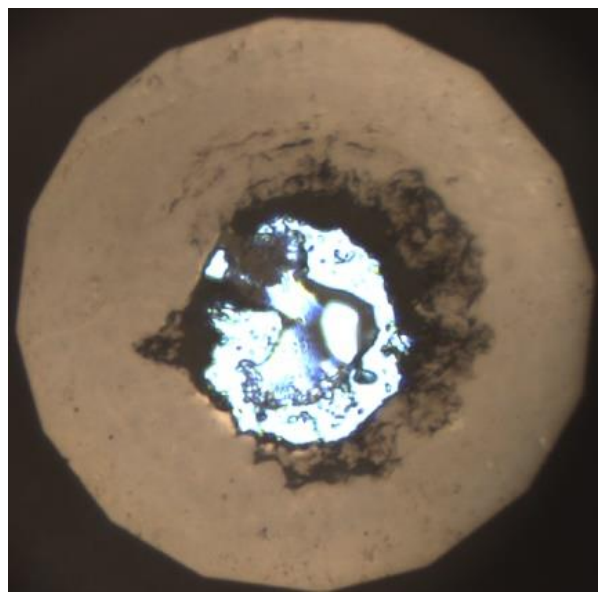

Figure S1. Sample chamber with methane crystals surrounded by He medium.

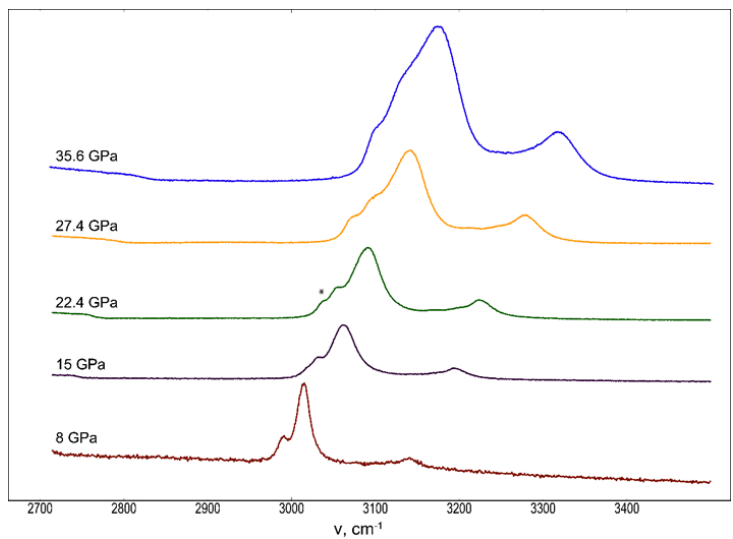

**Figure S2**. **Raman spectrum of methane at various pressures.** Methane was grown in phase B, identified via splitting of the symmetric C-H stretch mode $\nu_1$ at 3010 cm$^{-1}$. Methane is in phase B at 8 and 15 GPa and transforms to HP1 phase at higher pressures. Asterisk marks the appearance of the $\nu_1$ further peak splitting, manifesting the transition to HP1 phase. The collection time is 10 s and the laser power is 5 mW except the spectrum measures at 8 GPa, where 0.5 mW laser power was used. The excitation wavelength is 488 nm for 8 GPa.

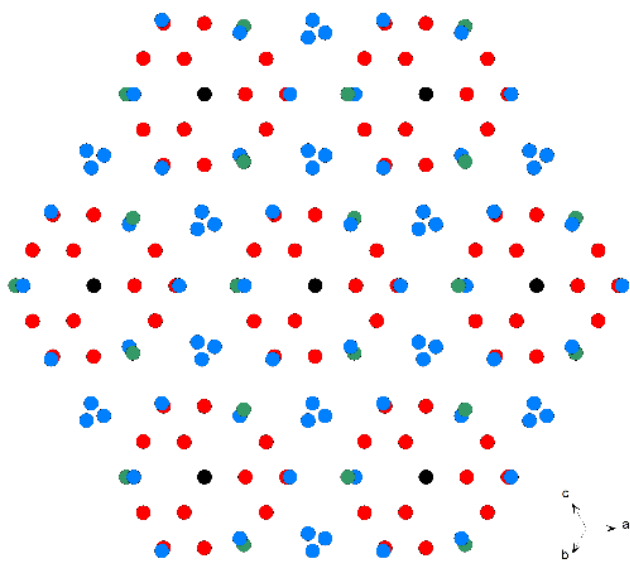

**Figure S3**. Crystal structure of methane-B. Black, red, green and blue balls show the positions of C1, C2, C3 and C4 atoms respectively (Table S1).

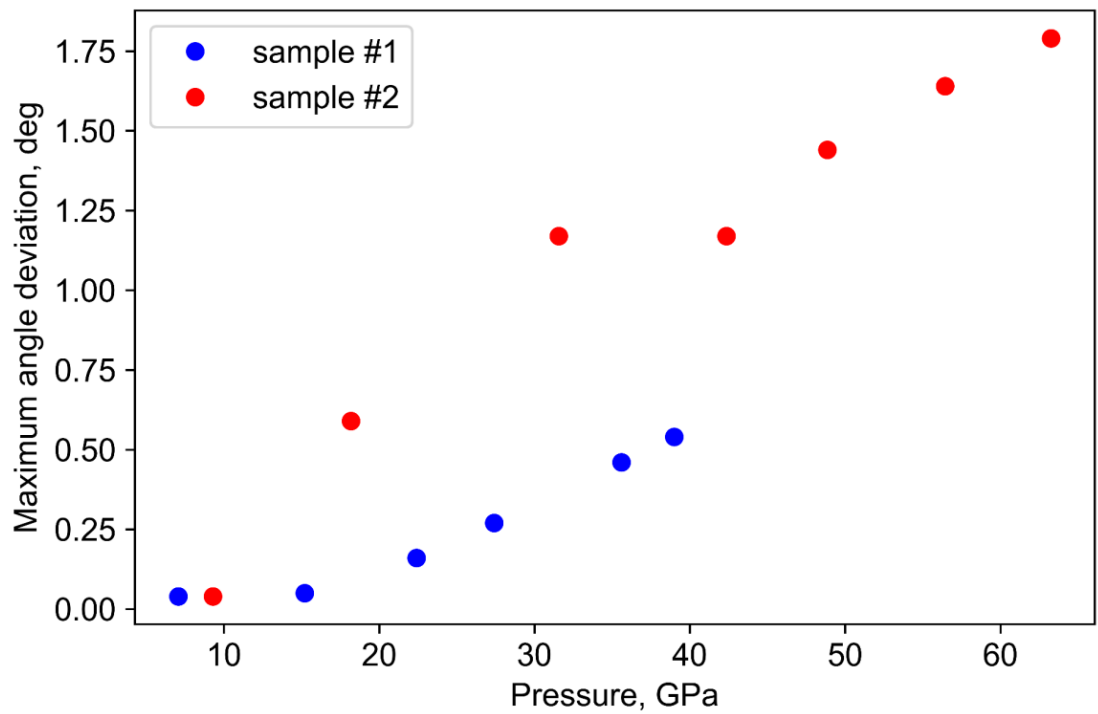

**Figure S4**. Maximum angle deviation from the ideal angles of the hexagonal lattice (α = β = 90°, γ = 120°).

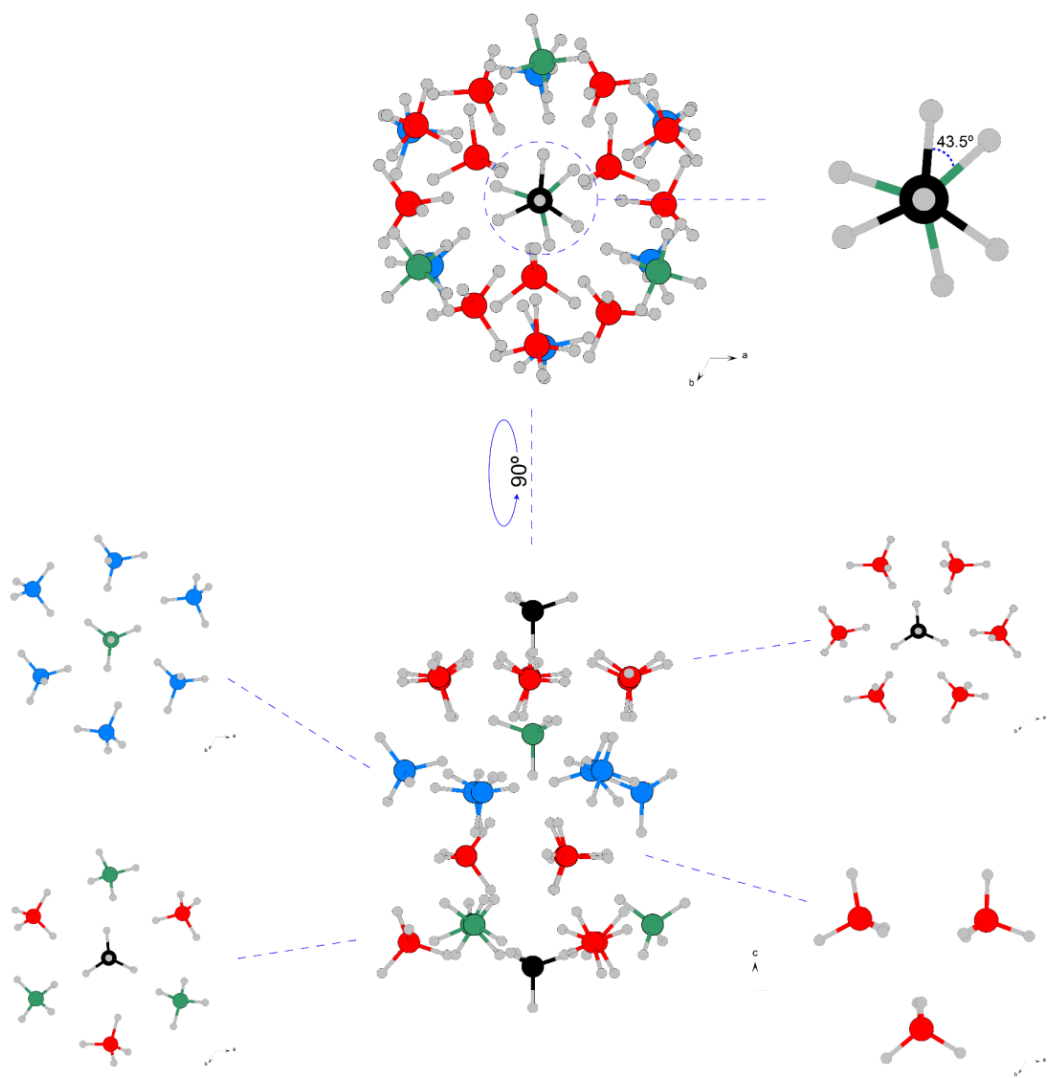

**Figure S5**. Crystal structure of HP CH$_4$ at 30 GPa. Black, red, green and blue balls represent the positions of C1, C2, C3 and C4 atoms respectively (Table S4).

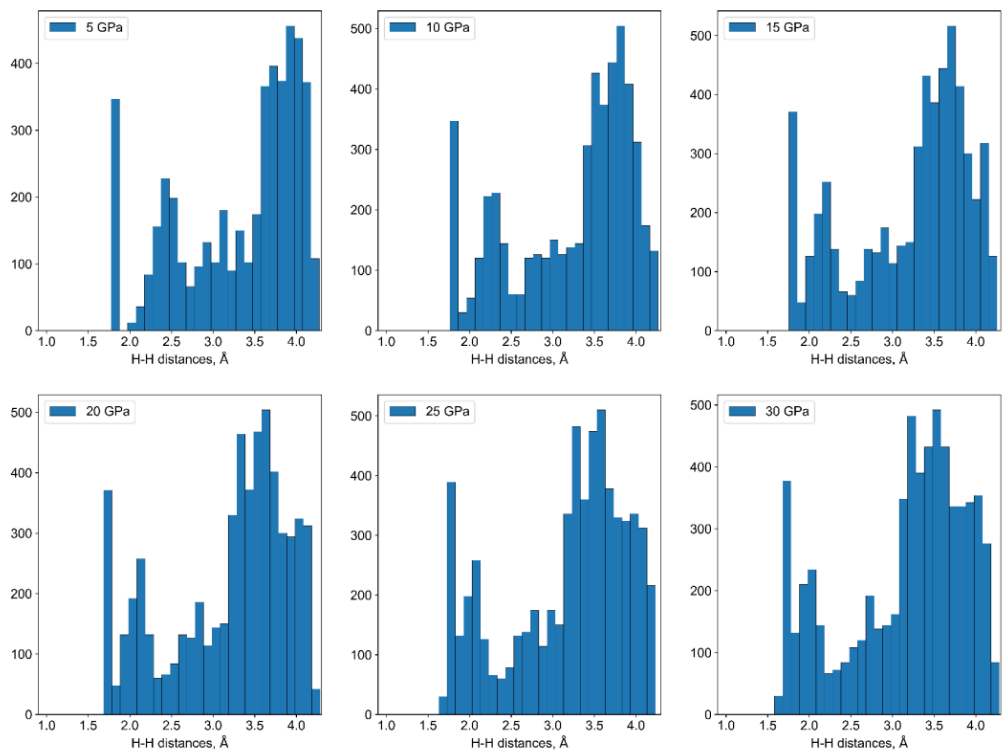

**Figure S6**. Distribution of H-H distances in calculated $CH_4$ HP structures at various pressures. Distances are given in Å, vertical axis represents the number of occurencies.

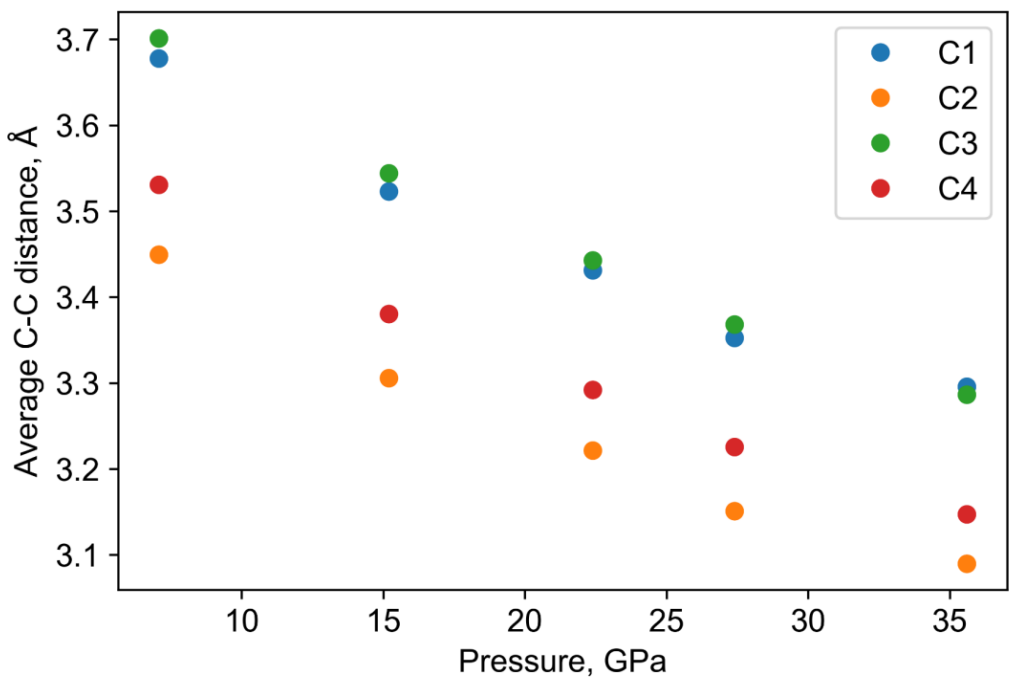

**Figure S7.** Average C-C distances of C1-C4 methane molecules to the next neighbors as a function of pressure.

**Table S1**. Crystal data, refinement and crystal structure details of Methane-B phase

| Pressure (GPa) | 7.07 | 15.2 |
|---|---|---|
| **Crystal data** | | |
| Crystal system, space group | Cubic, $I\bar{4}3m$ | |
| Temperature (K) | 293 | |
| $a$ (Å) | 11.942 (2) | 11.439 (2) |
| $V$ (Å$^3$) | 1703.0 (5) | 1496.9 (5) |
| $Z$ | 58 | |
| Radiation type | X-ray, $\lambda = 0.34453$ Å | |
| $\mu$ (mm$^{-1}$) | 0.03 | |
| **Data collection** | | |
| Diffractometer | GP @ ID16-B | |
| Absorption correction | Multi-scan | |
| No. of measured, independent and observed [$I > 3\sigma(I)$] reflections | 818, 323, 232 | 728, 279, 186 |
| $R_{int}$ | 0.072 | 0.054 |
| $(\sin \theta/\lambda)_{max}$ (Å$^{-1}$) | 0.756 | 0.744 |
| **Refinement** | | |
| $R[F^2 > 2\sigma(F^2)]$, $wR(F^2)$, $S$ | 0.106, 0.126, 3.16 | 0.108, 0.124, 3.52 |
| No. of reflections | 323 | 279 |
| No. of parameters | 17 | 17 |
| $\Delta\rho_{max}$, $\Delta\rho_{min}$ (e Å$^{-3}$) | 0.54, -0.60 | 0.31, -0.32 |
| **Crystal structure** | | |
| Atomic coordinates and Wyckoff positions | C1 - 2$a$ (0 0 0)<br>C2 - 24$g$ (0.0895(3) 0.0895(3) 0.2801(4))<br>C3 - 8$c$ (0.3211(4) 0.3211(4) 0.3211(4))<br>C4 - 24$g$ (0.1443(3) 0.1443(3) 0.5368(4)) | C1 - 2$a$ (0 0 0)<br>C2 – 24$g$ (0.0893(3) 0.0893(3) 0.2808(5))<br>C3 – 8$c$ (0.3216(4) 0.3216(4) 0.3216(4))<br>C4 – 24$g$ (0.1456(3) 0.1456(3) 0.5354(5)) |
| Atomic displacement parameters, $U_{iso}$(Å$^2$) | C1 - 0.031(2)<br>C2 - 0.0346(17)<br>C3 - 0.0331(14)<br>C4 - 0.0365(17) | C1 – 0.027(3)<br>C2 – 0.032 (2)<br>C3 – 0.0256(15)<br>C4 – 0.035(2) |